\documentclass[aps,prl,showpacs,twocolumn]{revtex4}

\usepackage[dvips]{graphicx}
\usepackage{epsfig}
\usepackage{amsmath}
\usepackage{amsfonts}
\usepackage{amssymb}
\usepackage{wasysym}

\expandafter\ifx\csname package@font\endcsname\relax\else
 \expandafter\expandafter
 \expandafter\usepackage
 \expandafter\expandafter
  \expandafter{\csname package@font\endcsname}%
\fi

\begin{document}
\def\be{\begin{equation}}
\def\ee{\end{equation}}

\title{The role of static stress diffusion in the spatio-temporal
organization of aftershocks }

\author{E. Lippiello$^{(1)}$, L. de Arcangelis$^{(2)}$, C. Godano$^{(1)}$}

\affiliation{ $^{(1)}$
Dept. Environmental Sciences, Second University of Naples and CNISM, 81100 Caserta, Italy\\
$^{(2)}$ 
Institute for Building Materials, ETH H\"onggerberg, 8093 Z\"urich, Switzerland,\\ 
Dept. Information Engineering, Second University of Naples and CNISM,
81031 Aversa (CE),
 Italy}

\begin{abstract}
We investigate
the spatial distribution of aftershocks 
%as function of space and time distances from
%the main-shock
 and  we find that aftershock linear density exhibits a
maximum,  that depends on the mainshock magnitude, followed by a power
law decay. The exponent controlling the asymptotic decay and
the fractal dimensionality of epicenters clearly indicate triggering
by static stress. The non monotonic behavior of the linear density and
its dependence on the mainshock magnitude can be interpreted in terms of
diffusion of static stress. This is supported by the power law
growth with exponent $H\simeq 0.5$ of the average main-aftershock distance. 
%In order to support the above conclusions we 
Implementing static stress diffusion  
within a stochastic model for aftershock occurrence we are able to  
%Numerical results very well reproduce}  
reproduce aftershock linear density  spatial decay, its dependence on 
the mainshock magnitude and its evolution in time.
%The coefficient of the diffusion process
%is consistent with the value obtained from simulations of three dimensional
%viscoelastic relaxation. 
\end{abstract}
\maketitle

Large earthquakes give rise to a sudden increase of
the seismic rate in the surrounding area. Aftershocks are often
observed  where mainshocks have  increased the 
static Coulomb stress \cite{rea,kin,har,wys} and their rate decays in
time in agreement with state-rate friction laws \cite{die,ste}.  
Aftershocks also occur in regions of reduced 
static stress \cite{par} as well as at  distances up to thousand
kms from the mainshock \cite{hil,sta,bro,gom,ebe,gom2}.   
Dynamic stress related to the passage of shock waves, is the most
plausible explanation for this remote triggering. 
Many studies, also supported by experiments on laboratory fault
gouge systems \cite{joh}, have recently proposed dynamic stress as the main 
mechanism responsible for aftershock triggering \cite{joh,kil,kil2,fel}. 
The distribution $\rho(\Delta r)$, where $\Delta r$ is the epicentral
distance between each aftershock and its related mainshock, 
 represents a useful 
tool to discriminate between triggering by static or dynamic
stress \cite{fel}. In both cases, $\rho(\Delta r)$ is expected to
decay asymptotically   
as $\Delta r^{-\mu}$, where 
 $\mu$ is related to the fractal dimensionality $D$ of
epicenters via the relationship $\mu+D-1=\alpha$ with $\alpha=1$ or
$\alpha=2$ for dynamic or static stress triggering, respectively.
Felzer \& Brodsky (FB) \cite{fel} studied $\rho(\Delta r)$ 
for small and intermediate mainshock
magnitudes, obtaining a pure power law decay with an exponent  $\mu
\simeq 1.4$. This result, together with the estimate
$D \simeq 1$, was interpreted in favor of  
dynamic stress triggering aftershocks. 
In this paper we will show that the distribution $\rho(\Delta r)$
exhibits a
non-monotonic behavior, with a power law tail and a 
maximum depending on the mainshock
magnitude that can be attributed to a stress diffusion mechanism.
% Furthermore the tail is strongly affected by
%non-correlated background events and therefore
%a power law fit is a subtle point. By properly taking into account
%background seismicity, we obtain a decay
%of $\rho(\Delta r)$ that supports the static stress
%scenario. Moreover, we relate the dependence on the mainshock magnitude
%to a stress diffusion mechanism.   

In our analysis we use the Shearer et
al. relocated Southern California Catalogue in the years
1981-2005 \cite{she} with an average uncertainty on the epicentral
localization of $0.03$ km. We consider all events with magnitude $m\ge2$. 
Mainshocks are identified with the same 
criterion used by FB, i.e  mainshocks are events separated in time and
space from larger earthquakes \cite{fel}. 
Aftershocks are all subsequent events occurring within a circular region of
radius $100$ km centered at the mainshock epicenter. 
In Fig.1 we plot
$\rho(\Delta r)$ for all aftershocks related to a mainshock with
magnitude $m \in [M,M+1[$ for $M=2,3,4$ and for a typical time window of
30 min post-mainshock, as considered by FB.
We find that $\rho(\Delta r)$ exhibits a maximum at a value of $\Delta r$
increasing with $M$, followed by a pure power law decay 
$\Delta r^{-1.9}$ only when $M=4$. For $M=2,3$, conversely,  a plateau is observed 
at large distances, $\Delta r >10 km$ ($M=2$) and $\Delta r>30 km$ 
($M=3$), which is related to uncorrelated background events.   
Indeed, $\rho(\Delta r)$  can be written as the
sum $\rho(\Delta r)=\rho_{AS}(\Delta r)+\rho_{B}(\Delta r)$,
where  $\rho_{AS}(\Delta r)$ is the aftershock density distribution and 
$\rho_B(\Delta r) \propto \Delta r^{D-1}$ is the contribution
of background events. 
% The latter contribution  
%can be evidenced by evaluating the $\rho(\Delta r)$ only for events with occurrence 
%times very distant from the mainshock.} 
Since the aftershock number decreases in time 
whereas background seismicity has a constant
rate,  $\rho(\Delta r) \simeq \rho_B(\Delta
r)$ in temporal windows sufficiently distant from the mainshock.  
More precisely, we obtain 
$\rho_B(\Delta r)$ in temporal widows distant more than
$t_d=70$ days  
from the mainshock.
Results, plotted as open symbols in Fig.1,  do not depend on $t_d$ for
larger $t_d$.
For each $M$, a flat behaviour is obtained for $\Delta r>1$ km,
implying $D \simeq 1$, in agreement with FB. 
A more precise measurement gives $D=1.03 \pm 0.05$. 
%The initial decay of $\rho_B(\Delta r)$ for $M=2$ (Fig.1a) can be
%attributed to the finite probability that, even if the temporal window of $30$ 
%min is randomly chosen,  
%an aftershock sequence is present in the interval. This probability
%becomes vanishing for larger $M$.  
The value of $\rho_B(\Delta r)$ depends
on $M$, since it is proportional to the number of mainshocks in each
class $M$. This implies that $\rho_B(\Delta r)$ becomes less relevant for
larger $M$ and, in particular, does not affect the exponent 
$\mu=1.88 \pm 0.05$ obtained for $M=4$ from  Fig.1. For $M=2,3$,
conversely, the tail of the distribution  must be 
appropriately  fitted with  $\rho(\Delta
r)=\rho_B (\Delta r)+ A \Delta r^{-\mu}$. For $\Delta r>1$ km, the
correlation coefficient provides results consistent
with $\mu=2$ and excludes  $\mu=1.4$. Hence, the exponent value $\mu  \simeq
1.4$ obtained as
best fit in the range $[0.2:50]$ km (orange line in Fig. 1b)
does not represent the asymptotic decay of $\rho(\Delta r)$.
Similar behavior is obtained for hypocentral distances, with small
differences only at lengths comparable with location errors.

In order to extend the analysis to larger temporal windows
post-mainshock we use the         
criterion proposed in ref. \cite{bai} to separate aftershocks from
background events. 
%In particular, 
%we consider a two dimensional grid of lattice spacing $\Theta=0.5^o$ covering
%the entire Southern California. For each site $\vec x$ on the
%grid, we compute the number of $m \ge 3$ earthquakes occurred in
%$T_0=1$ year inside a circle of radius $\Delta=50$ km centered at
%$\vec x$. The mean number of events per year $n(\vec x)$ is obtained
%by averaging over the years covered by the catalog. 
Given two
events with magnitude $m_1$ and $m_2$ with occurrence times $t_1<t_2$
and locations $\vec r_1,\vec r_2$, the expected number of events inside
a circle of radius $\Delta r=\vert \vec r_1 -\vec r_2 \vert$ centered in
$\vec r_1$, over a time window $T=t_2-t_1$ is proportional to 
$n_{exp}(1,2)=C 10^{-b (m_1-2)} T \Delta r^{D}$. Here $D=1.03$,
$b$ is the slope of 
the Gutenberg-Richter magnitude-frequency distribution and
$C=2.06\ \ 10^{-11} sec^{-1}km^{-D}$ 
is the average rate of $m\ge2$ earthquakes in the catalog. 
For a given mainshock $(\vec r_1,t_1)$ each subsequent earthquake
($\vec r_2,t_2$) with 
$n_{exp}(1,2) < n_{th} \ll 1$, where $n_{th}$ is a given
threshold, is highly unexpected and therefore it is considered an aftershock. 
Aftershock number should decay in time according to the Omori law, which 
fixes the value of the threshold $n_{th}$, in particular we find $n_{th}=10^{-3}$. 
Different values of $D \in[1.1,1.6]$ provide similar results. 
This criterion allows to discriminate between 
aftershocks directly triggered by the mainshock (first generation)
from higher order generations, excluding eventual effects due to
aftershock cascading \cite{hel,huc,mck,mar}.    
An event $2$ is a first generation aftershock of the event $1$, if
in the time interval $]t_1,t_2[$ no event $j$
with $n(j,2)\le n(1,2)$ is present. All the following results are
obtained considering only first generation aftershocks. No important
difference is observed if higher order generation aftershocks are
included in the analysis.  
% by computing $n_{exp}(1,2)$ for all possible
%couples of events. The couple of events with the minimum
%$n_{exp}(1,2)$ is the most unexpected and therefore event $(\vec
%r_2,t_2)$ is triggered by event $(\vec r_1,t_1)$. If event $(\vec
%r_1,t_1)$ is a mainshock, event $(\vec r_2,t_2)$ is a first generation
%aftershock.                      
The study of  $\rho_{AS}(\Delta r)$ with this aftershock selection criterion
(Fig.2) provides results in agreement with the previous analysis,
i.e. a power law decay  
with an exponent $\mu \simeq 2$ for all values of $M$. 
Furthermore, curves for different $M$  collapse on the same
master curve (inset a of Fig.2) following   the scaling  
\be
\rho(\Delta r)=10^{-\beta M}F\left(\frac{\Delta r}{10^{\beta M}} \right ) 
\ee
with $\beta=0.42 \pm 0.02$. This result was obtained in ref. \cite{bai}
using a different  mainshock selection criterion. 
The function $F(x)$ is non monotonic and
exhibits power law behaviour $F(x) \sim x^{-\mu}$ with $\mu=1.94 \pm
0.04$ at large $x$. 
 The collapse of curves with small $M$ on curves with larger
$M$, weakly affected by the background seismicity, validates
the aftershock selection criterion.
%The same analysis for the Northern California Catalog provides similar
%results. 
Fig.2 confirms $\mu \simeq 2$ supporting the
static stress triggering scenario.

\begin{figure}
  \includegraphics[width=8cm]{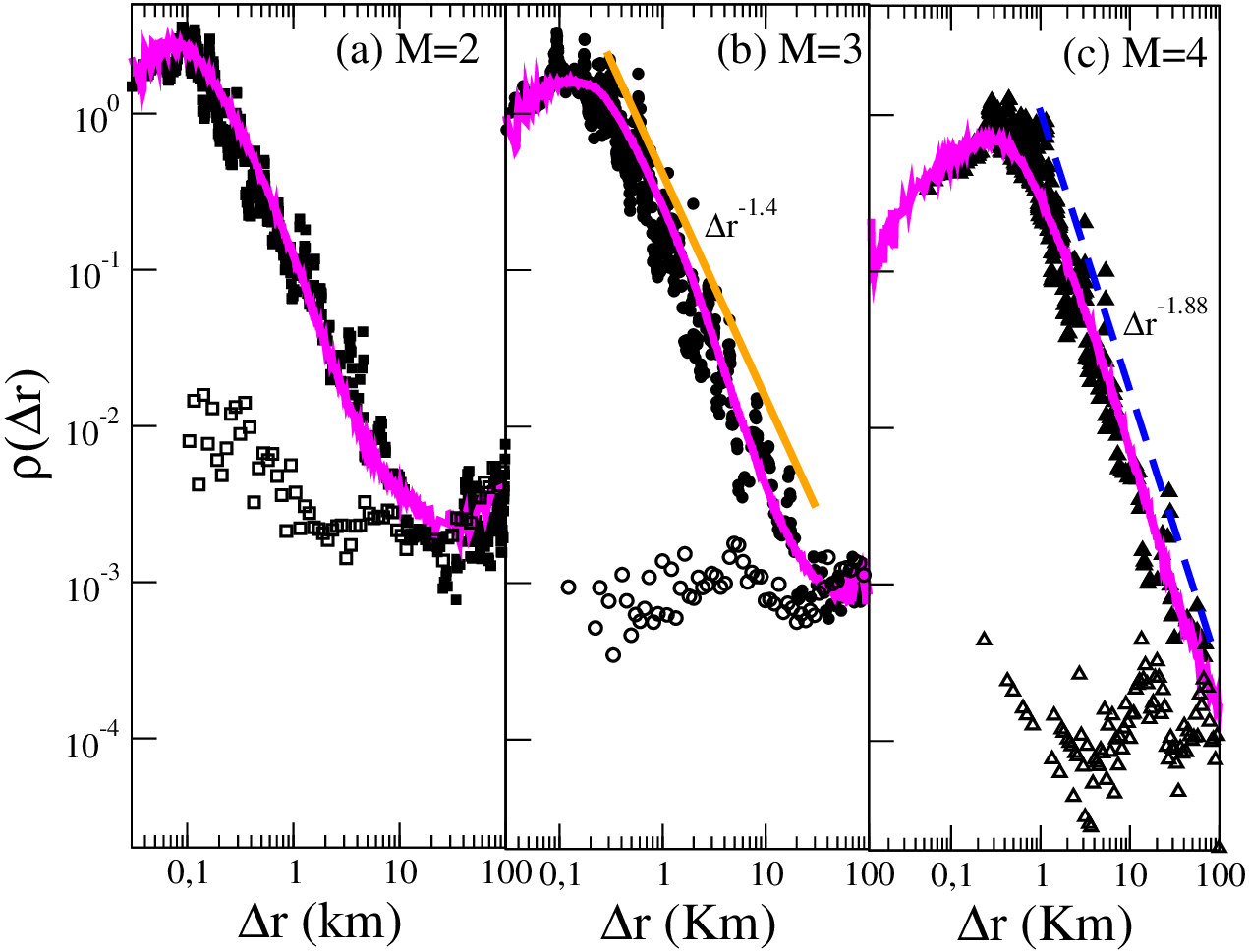}% Here is how to import EPS art
\caption{(Color online) 
The distribution of distances from the mainshock  (filled
  symbols) versus $\Delta r$ for
mainshock magnitude $m\in [M,M+1[$. 
Aftershocks are events occurring within $T=30 min$ from the mainshock 
($678, 864, 494$ aftershocks for $M=2, 3, 4$ respectively). 
Open symbols represent $\rho_B(\Delta r)$.  For $M=4$,  
the power law fit in the range
  $[1:100]$ km gives $\mu =1.88 \pm 0.05$ 
  (dashed blue line in panel c). 
Magenta curves are obtained by adding the
  experimental $\rho_B(\Delta r)$ to the numerical
  $\rho(\Delta r)$. The orange line in panel (b) is the power law  
  $x^{-1.4}$ obtained by FB in the intermediate range $[0.2:16]$km.}
\end{figure} 

\begin{figure}
\includegraphics[width=8cm]{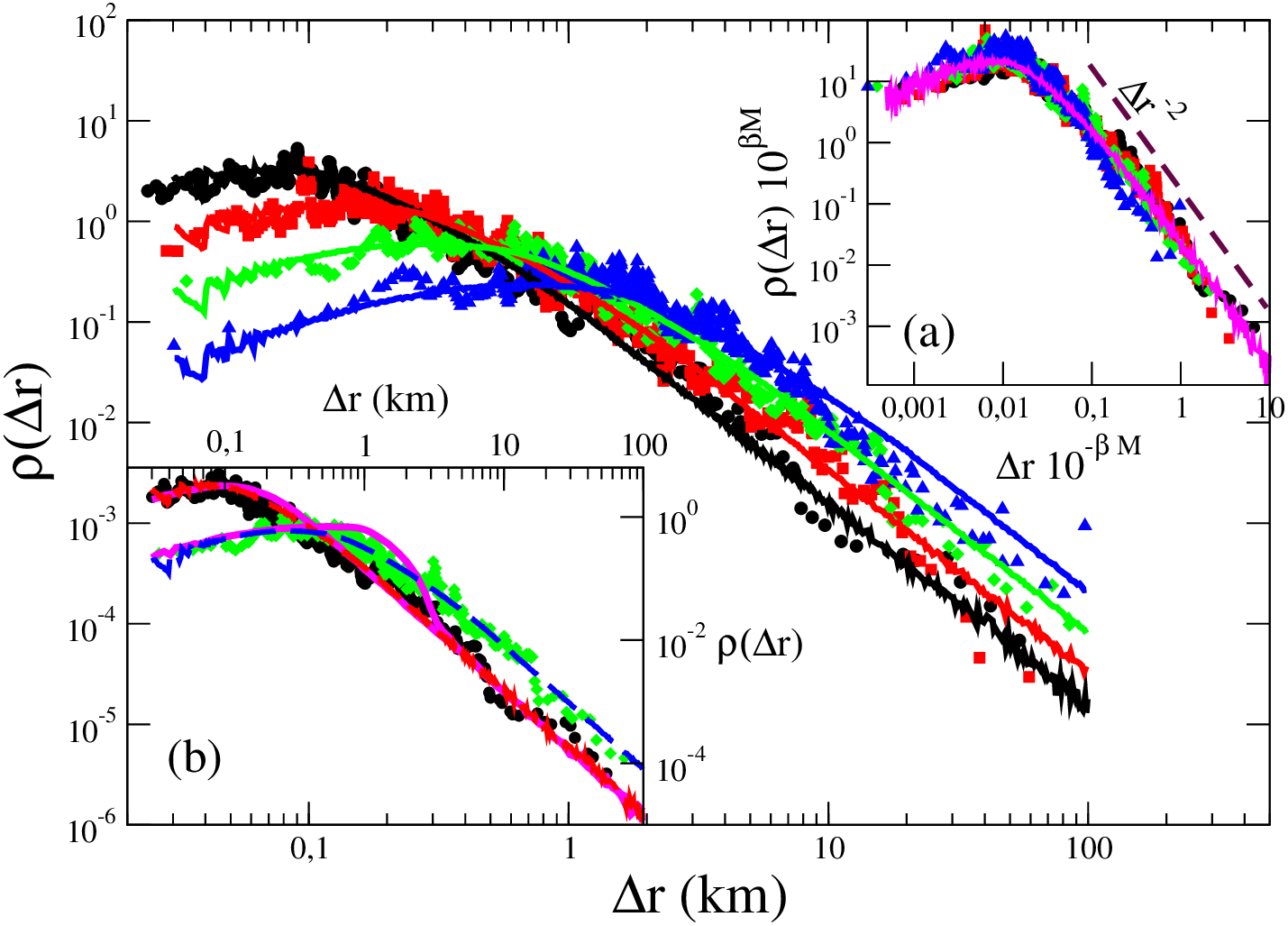}% Here is how to import EPS art
\caption{(Color online) 
The distribution of distances from the mainshock for $M=2$ (circles), 
$M=3$ (squares),
$M=4$ (diamonds) and $M=5$ (triangles). Aftershocks are events occurring
within $T=5h$ from the mainshock. 
%Results do not appreciably  depend on $T$. 
The aftershock (mainshock) number is $1065$ ($12746$) for $M=2$, 
$1800$ ($3410$) for $M=3$, $1425$ ($349$) for $M=4$ and 
$1454$ ($52$) for $M=5$.
Continuous lines are the result of numerical simulations. 
In the inset (a), collapse of the curve is obtained rescaling $\Delta
  r$ by  $10^{\beta M}$ according to Eq.(1), with
$\beta=0.42$. The continuous magenta line is the theoretical master
  curve $F(x)$  and the brown dashed line  shows the asymptotic decay
  $F(x) \sim x^{-2}$.
 In the inset (b), comparison of experimental $\rho(\Delta r)$ for $M=2,4$
(symbols) with the theoretical predictions $\rho_{th}(\Delta r)$
(continuous magenta lines). Dashed lines (red $M=2$, blue $M=4$) 
are the results of the stochastic model simulations. } 
\end{figure}

The non-monotonic behaviour of $\rho(\Delta r)$ is commonly attributed
to the violation of the point-source hypothesis \cite{fel}. This
implies that seismic sources have a
finite extension whose linear size scales with the
earthquake magnitude $L_S(m)=0.01 \ \ 10^{0.5m}$ km
\cite{Kagan}.  
One then computes $\rho_{th}(\Delta r)$ 
assuming that aftershocks are distributed
according to a power law from a point randomly chosen on the mainshock
fault and 
defining $\Delta r$ as the distance from the center of the
mainshock fault.
$\rho_{th}(\Delta r)$  follows the experimental $\rho(\Delta r)$
in the whole spatial range for $M=2$ (inset (b) in Fig.2). 
For larger $M$, conversely,
theoretical curves significantly deviates from the experimental ones.
Indeed, curves for different $M$ collapse on the same pure power law
decay  at distances $\Delta r>L_S(m)$,  where the point source hypothesis
holds. This implies that, even if theoretical curves exhibit a
non-monotonic behavior, they do not verify the scaling collapse
Eq(1).

The  scaling behavior of $\rho(\Delta r)$ can be
attributed to a diffusion process.
 To this extent, we implement  static stress diffusion in a stochastic 
model for seismic occurrence based on a dynamical scaling assumption 
\cite{lip,lip2,lip3}. Within this framework, for a given
mainshock of magnitude $m_0$ and an aftershock of magnitude $m$, the magnitude 
difference $\Delta m = m_0-m$, $\Delta r$ and $\Delta t$ are not independent 
variables.
More precisely, if time is rescaled by a a generic scaling factor $\lambda$, 
$\Delta t\to \lambda\Delta t$, the statistical properties are invariant provided that
$\Delta r \to \lambda^{H} \Delta r$ and  $\Delta m\to \Delta m +
(1/b)\log\lambda$, where $H$ is a  scaling exponent. 
The scaling relation among $\Delta t$, $\Delta r$ and $\Delta
m$ implies that,  for a given
mainshock of magnitude $m_0$, the conditional probability 
to have a magnitude $m$
aftershock at distance $\Delta r$ after a time  $\Delta t$, takes the scaling
form
$P(\Delta t,\Delta r,m,m_0)=\Delta t^{-H} G_t\left (\frac{10^{b
(m_0-m)}}{\Delta t}\right) 
G_r\left (\frac{\Delta r}{\Delta t^{H}}\right)$.
Under the only assumption that $G_t(y)$ and $G_r(x)$ are normalizable
functions, one recovers several features of seismic occurrence as the
GR law, the generalized Omori law, the scaling behavior of the
intertime distribution \cite{lip}. 
The distribution $\rho(\Delta r)$  can be 
obtained by integrating $P(\Delta t,\Delta r,m,m_0) P(m_0)$ over $\Delta t$ and $m$. The scaling relation for $P(\Delta t,\Delta
r,m,m_0)$ and the GR law $P(m_0) \sim 10^{-b m_0}$
then give Eq.(1), with $F(x) \propto \int_0^{\infty}
du\int_0^{\infty} dv u^{-1}v^{-H} G_t(u/v) G_r(x v^{-H})$ and
$\beta=b H$. 
Assuming the power law decay
$G_r(x) \propto x^{-\mu}$, for $x$ larger than a cut-off $x_0$, 
$F(x)$ is a non-monotonic function with an asymptotic  
decay $F(x) \sim x^{-\mu}$  for  $x \gg 1$. We therefore implement 
in the numerical simulations the parameters fitted from experimental data,
$\mu=2$ and $H=0.47$,  obtained from $\beta=0.42$ and the typical value $b=0.9$.
In particular, following the procedure described in \cite{lip2}, we set   
$G_t(y) \propto \left(e^{1/(k_t y)} -1 +\gamma_1\right)^{-1}$
with the parameters
$k_t=12.7 h$, $\gamma_1=0.1$, $b=0.9$, and $G_r(x) \propto (k_r
x)^{-\mu}$ for $x > x_0$  with $\mu=2$, $H=0.48$, $k_r=5.1\ \ 10^{-6}
km^z/sec $ and $x_0=3\ \  10^{-3} km /sec^z$.
We find that, for all values of $M$, numerical curves follow the experimental
ones (Fig.2). The scaling (1) is  fulfilled with the numerical
$F(x)$ reproducing the experimental master curve (inset (a) of
Fig.2). As a further check, we add  
$\rho_B(\Delta r)$, obtained in Fig.1, to the numerical distribution
$\rho(\Delta r)$. Numerical results (Fig.1) very well agree
with experimental data over the entire spatial range. 

\begin{figure}
\includegraphics[width=8cm]{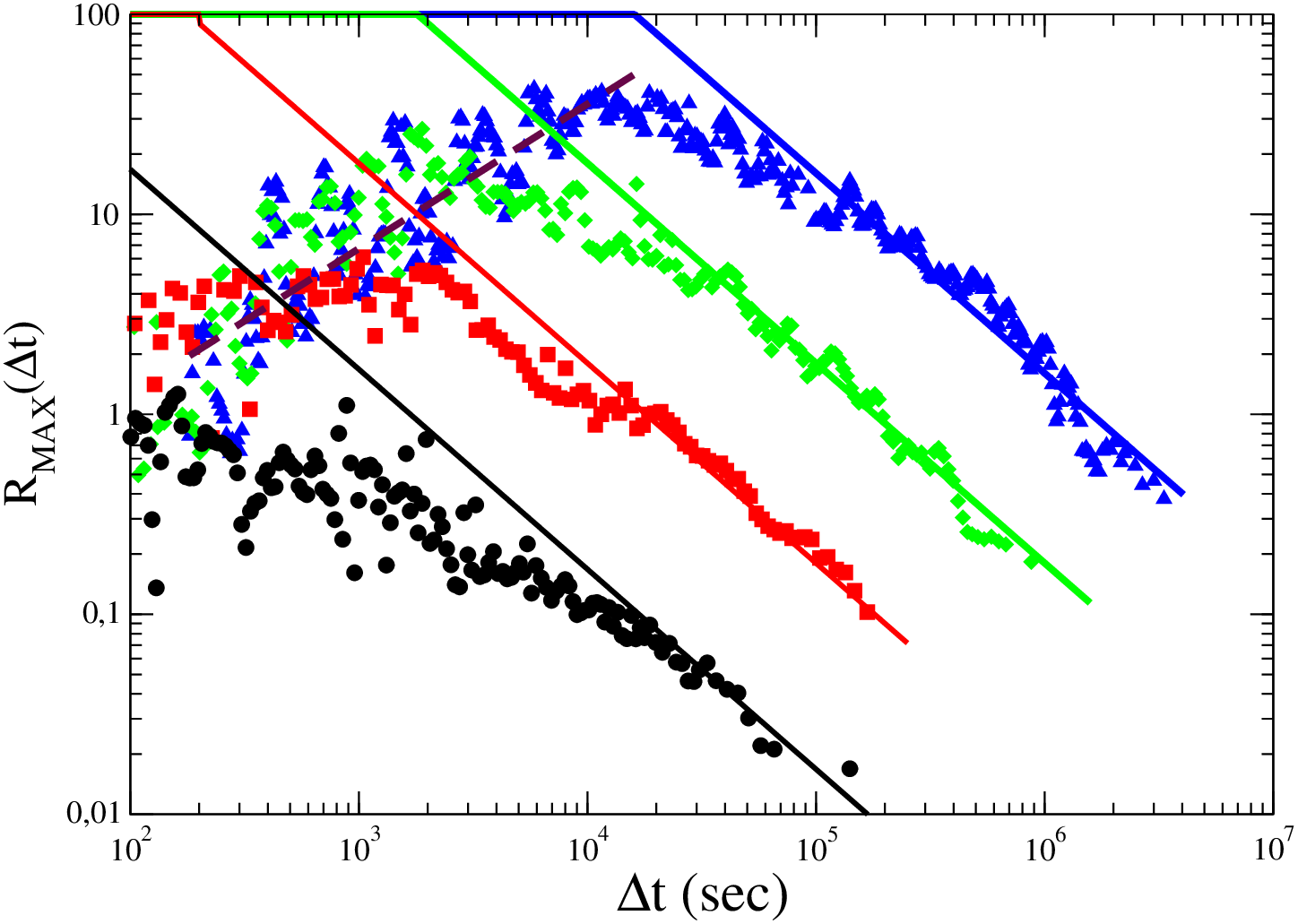}% Here is how to import EPS art
\caption{(Color online) 
The maximum distance $R_{MAX}(\Delta t)$ for $M=2,3,4,5$ (circles, squares, 
diamonds, triangles) from bottom to top. 
$R_{MAX}(\Delta t)$ grows until $\Delta t_M$ ($\Delta t_M=16000
 sec, 2900 sec, 690 sec, 190 sec$ for $M=5,4,3,2$). The dashed line is
 the power law fit $R_{MAX}(\Delta t) \sim \Delta t^H$  with
 $H=0.54\pm0.05$ obtained for $M=5$ and $\Delta t<\Delta t_M$. 
Continuous lines are the theoretical prediction for $L_{MAX}(\Delta t)$.}
\end{figure}
 
The agreement between experimental and numerical results 
supports the validity of the scaling relation
 $\Delta r \sim \Delta t^{H}$ with $H \simeq 0.5$, which implies that 
the evolution in time of stress 
%variation, produced by the mainshock, 
is consistent with a diffusion equation. More direct evidence of static stress
diffusion can be obtained by the temporal evolution of the
main-aftershock spatial distance. In particular we compute   
$R_{MAX}(\Delta t)$ ($R(\Delta t)$), i.e.   
the maximum (average) distance from a  mainshock with 
$m \in [M,M+1[$, of aftershocks occurring in  the  time window  
$[\Delta t,\Delta t(1+\epsilon)]$.
%In this analysis, we include only aftershocks directly triggered by the
%mainshock. 
For all $M$, $R_{MAX}(\Delta t)$ exhibits (Fig.3)
a non monotonic behaviour with a maximum at a $M$-dependent
typical $\Delta t_M$. $\Delta t_M$ can be identified as the time when
the percentage of events identified  as aftershocks becomes smaller
than  the $90\%$
of the total number of recorded earthquakes.  
Therefore, for  $\Delta t< \Delta t_M$, 
no significant bias
related to the aftershock selection procedure is present. 
In this temporal regime, similar results are obtained including
in the analysis 
all subsequent earthquakes occurring 
within a radius of $100$ kms from the mainshock.
Fig.3 shows that, for all values of $M$,   $R_{MAX}$ increases for
times $\Delta t< \Delta t_M$. For $M=5$ where $\Delta t_M= 16000 sec$,
a power law regime $R_{MAX}(\Delta t) \sim \Delta t^H$
clearly detected with $H=0.54\pm0.05$.
On the other hand,
the decay for $\Delta t>\Delta t_M$      
originates from a
bias introduced by the method for aftershock selection. 
The condition $n_{exp}(1,2)<n_{th}$, indeed, implies that
aftershocks are only events occurring within a given
temporal-magnitude region and, in particular, 
all events occurring at distances larger
than $L_{MAX}(\Delta t) \propto 10^{M b/D} (\Delta t)^{-1/D} $ are not
considered as aftershocks.
The tails of $R_{MAX}(\Delta t)$ are consistent with a pure power law 
decay $\Delta t^{-1/D}$ in agreement with  the 
analytical expression for $L_{MAX}(\Delta t)$ (Fig.3). 

Further indication of diffusion   can be obtained in the regime $\Delta t>\Delta
t_M$ by considering the average distance $R(\Delta t)$ inside a region of radius $L_{sup}$.
This can be evaluated as $R(\Delta t) = \int _{0}^{L_{sup}} d\Delta r \Delta r
\rho(\Delta r) /\int _{0}^{L_{sup}} d \Delta r \rho(\Delta r)$,
using the decay  $\rho(\Delta r) \propto (\Delta r+K)^{-2}$ obtained from Fig.2 
\be
R(\Delta t)= K \left [ \left ( 1+\frac{K}{L_{sup}} \right) 
 \log  \left (1+\frac {L_{sup}}{K} \right ) -1 \right ] .
\ee
According to the previous analysis $L_{sup}=100$ km when $\Delta
t<\Delta t_M$ and $L_{sup}=L_{MAX}(\Delta t)$ when $\Delta t>\Delta
t_M$. We introduce in the above equation $K=B \Delta t^{H}$ with 
$B=0.018 km/sec^H$ and $H=0.47$,    
obtained from the numerical simulations. 
Fig.4 shows that for all $M$, without any
further parameter tuning, the theoretical prediction (2) reproduces
experimental results in the whole time range. In Fig.4 we also 
plot Eq.2 assuming a constant $K$, obtained as the best fit from
Fig.2.
In this case, the theoretical $R(\Delta t)$ (dashed lines in Fig.4) overestimates the experimental 
$R(\Delta t)$ at small $\Delta t$, whereas it somehow
underestimates it at larger times.  
 Previous analyses \cite{hel,huc,mck,mar} have obtained a smaller value of the 
diffusion exponent, $H\simeq 0.1$. 
The basic differences with our study is that 
%and previous results are: 
%i) we have only considered aftershocks directly triggered by the
%mainshock, whereas aftershocks triggered by previous aftershocks are included in 
%other studies; ii) 
in previous analyses aftershocks have not been classified 
according
to the mainshock magnitude and distances significantly smaller than
the mainshock fault length have been included in the analysis. 
%also large mainshocks, that do not
%verify the point source hypothesis, have been considered. 
%This can
%give rise to a smaller value for $H$. 
Interestingly, McKernon and
Main \cite{mck}
recover $H\simeq 0.5$ at very large distances, where the point
source hypothesis is recovered.  

\begin{figure}
\includegraphics[width=8cm]{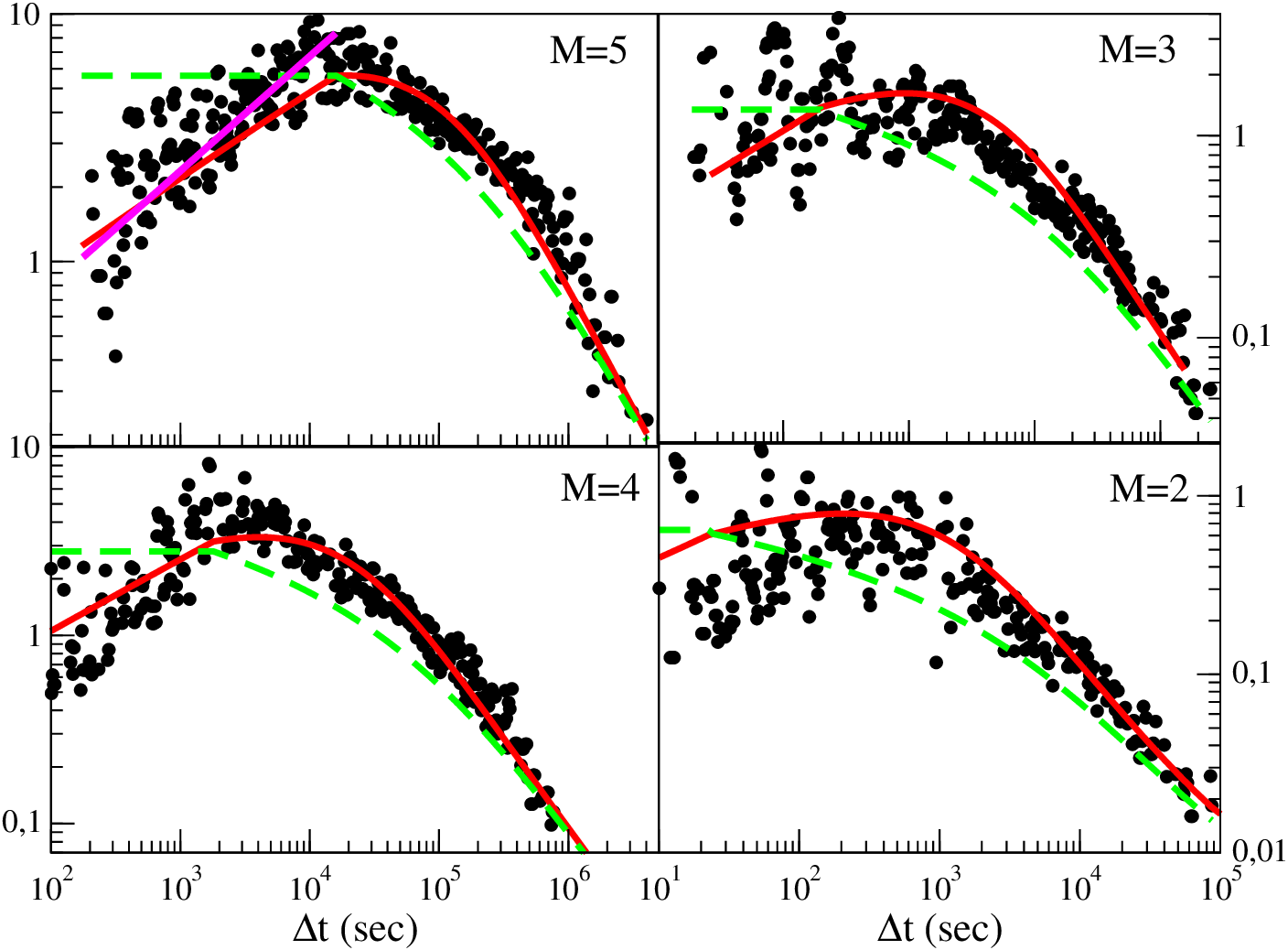}% Here is how to import EPS art
\caption{(Color online) 
The average spatial distance $R(\Delta t)$ from the mainshock (black circles)
of aftershocks occurring in the time window $[\Delta t,\Delta
  t(1+\epsilon)]$ with $\epsilon=0.02$   versus $\Delta t$.
The initial growth $R(\Delta t) \sim \Delta t^{0.47}$ (magenta
 line) is consistent with the diffusion behaviour. The decay
 at larger times is related to the upper cut-off $L_{sup}$.
Continuous red and dashed green curves are the theoretical prediction Eq.(2): 
% that takes into account $L_{sup}$ and uses $\rho(\Delta r) \sim
% (\Delta r+K)^{-\mu}$ with $\mu=2$ fitted from Fig.2.
Green curves correspond to 
 $K=1.8 \ \ 10^{0.42 (M-5)}$ km fitted from Fig.2. Red curves correspond to 
 $K=0.018 \ \ \Delta t^{0.47}$ km, where $\Delta t$ is measured in seconds,
 obtained from the numerical model. 
% Similar results are obtained for $D=1.6$.
}
\end{figure}

In conclusion, we have shown that static stress is the main mechanism
responsible for aftershock occurrence.  Indeed, by properly taking into
account background seismicity, $\rho (\Delta r)$ exhibits the scaling
behavior (1) with the power law decay expected within the static
stress triggering scenario. Moreover,  
the very good agreement  of the theoretical prediction (2) with the
 numerical results  and experimental data indicates that the 
aftershock spatial organization evolves in time according to a
diffusion equation. 
Migration of aftershocks \cite{mog} is often observed   
and interpreted within different contexts, including 
state/rate friction \cite{die,ste}, viscoelastic relaxation process \cite{ryd,fre,jon}
and aftershock cascading \cite{hel,huc,mck,mar}. 
In the present study, the latter mechanism can be
discarded, since only aftershocks directly triggered by the mainshock
have been considered. The estimated value $B=0.018 km/sec^H$ 
predicts, on average, a post seismic stress change over a region of
 about $10^2$ km in $7$ years. This
is consistent with simulations of $3d$ viscoelastic post
 seismic relaxation after the 1992 Landers earthquake \cite{fre}.

\end{document}